\documentclass{iaus}
\usepackage{graphicx}
\renewcommand{\theequation}{\arabic{equation}}

\title[Reference Frames and Transformations for Lunar Motion] 
{Reference Frames, Gauge Transformations and Gravitomagnetism in the Post-Newtonian Theory of the Lunar Motion}

\author[Yi Xie \& Sergei Kopeikin]   
{Yi Xie$^{1,2}$
\and Sergei Kopeikin$^2$\footnote{email: {\tt kopeikins@missouri.edu}}}

\affiliation{$^1$Astronomy Department, Nanjing University, \\ Nanjing, Jiangsu 210093, China \\[\affilskip]
$^2$Department of Physics \& Astronomy, University of Missouri-Columbia, \\ Columbia, MO 65211, USA}

\pubyear{2009}
\volume{261}  
\pagerange{1--5}
\jname{Relativity in Fundamental Astronomy: Dynamics, Reference Frames, and Data Analysis}
\editors{Sergei Klioner, P. Kenneth Seidelmann \& Michael Soffel, eds.}
\begin{document}

\maketitle

\begin{abstract}

We construct a set of reference frames for description of the orbital and rotational motion of the Moon. We use a scalar-tensor theory of gravity depending on two parameters of the parametrized post-Newtonian (PPN) formalism and utilize the concepts of the relativistic resolutions on reference frames adopted by the International Astronomical Union in 2000. We assume that the solar system is isolated and space-time is asymptotically flat. The primary reference frame has the origin at the solar-system barycenter (SSB) and spatial axes are going to infinity. The SSB frame is not rotating with respect to distant quasars. The secondary reference frame has the origin at the Earth-Moon barycenter (EMB). The EMB frame is local with its spatial axes spreading out to the orbits of Venus and Mars and not rotating dynamically in the sense that both the Coriolis and centripetal forces acting on a free-falling test particle, moving with respect to the EMB frame, are excluded. Two other local frames, the geocentric (GRF) and the selenocentric (SRF) frames, have the origin at the center of mass of the Earth and Moon respectively. They are both introduced in order to connect the coordinate description of the lunar motion, observer on the Earth, and a retro-reflector on the Moon to the observable quantities which are the proper time and the laser-ranging distance. We solve the gravity field equations and find the metric tensor and the scalar field in all frames. We also derive the post-Newtonian coordinate transformations between the frames and analyze the residual gauge freedom of the solutions of the field equations. We discuss the gravitomagnetic effects in the barycentric equations of the motion of the Moon and argue that they are beyond the current accuracy of lunar laser ranging (LLR) observations.
\end{abstract}


The tremendous progress in technology, which we have witnessed during the last 30 years, has led to enormous improvements of precision in measuring time and distances within the boundaries of the solar system. Observational techniques like lunar and satellite laser ranging, radar and Doppler ranging, very long baseline interferometry, high-precision atomic clocks, gyroscopes, etc. have made it possible to start probing the kinematic and dynamic effects in motion of celestial bodies to unprecedented level of fundamental interest. Current accuracy requirements make it inevitable to formulate the most critical astronomical data-processing procedures in the framework of Einstein's general theory of relativity. This is because major relativistic effects are several orders of magnitude larger than the technical threshold of practical observations and in order to interpret the results of such observations, one has to build physically-adequate relativistic models.  The future projects will require introduction of higher-order relativistic models supplemented with the corresponding parametrization of the relativistic effects, which will affect the observations.

The dynamical modeling for the solar system (major and minor planets), for deep space navigation, and for the dynamics of Earth's satellites and the Moon must be consistent with general relativity. LLR measurements are particularly crucial for testing general relativistic predictions and advanced exploration of other laws of fundamental gravitational physics. Current LLR technologies allow us to arrange the measurement of the distance from a laser on the Earth to a corner-cube reflector (CCR) on the Moon with a precision approaching 1 millimeter (\cite[Battat \etal\ 2007]{battat} and \cite[Murphy \etal\ 2008]{murthyetal08}).

At this precision, the LLR model must take into account all the classical and relativistic effects in the orbital and rotational motion of the Moon and Earth. Although a lot of effort has been made in constructing this model, there are still many controversial issues, which obscure the progress in better understanding of the fundamental principles of the relativistic model of the Earth-Moon system. 

The theoretical approach used for construction of the JPL ephemeris accepts that the post-Newtonian description of the planetary motions can be achieved with the Einstein-Infeld-Hoffmann (EIH) equations of motion of point-like masses (\cite[Einstein \etal\ 1938]{eih}), which have been  independently derived by \cite[Petrova (1949)]{petrova} and \cite[Fock (1959)]{fockbook} for massive fluid balls as well as by \cite[Lorentz \& Droste (1917)]{lord} under assumptions that the bodies are spherical, homogeneous and consist of incompressible fluid. These relativistic equations are valid in the barycentric frame of the solar system with time coordinate $t$ and spatial coordinates $x^i\equiv{\mathbf{x}}$.

However, due to the covariant nature of the general theory of relativity the barycentric coordinates are not unique and are defined up to the space-time transformation (\cite[Brumberg 1972]{vab}, \cite[Brumberg 1991]{brum}, and \cite[Soffel 1989]{sof89})
\begin{eqnarray}
	\label{gt1}
	t & \mapsto & t-\frac{1}{c^4}\sum_{\tiny{B}}\nu_{\tiny{B}}\frac{G\mathrm{M}_{\tiny{B}}}{R_{\tiny{B}}}(\mathbf{R}_{\tiny{B}}\cdot \mathbf{v}_{\tiny{B}})\;,\\
	\label{gt2}
	\mathbf{x} & \mapsto & \mathbf{x}-\frac{1}{c^2}\sum_{\tiny{B}}\lambda_{\tiny{B}}\frac{G\mathrm{M}_{\tiny{B}}}{R_{\tiny{B}}}\mathbf{R}_{\tiny{B}}\;,
\end{eqnarray}
where summation goes over all the massive bodies of the solar system ($B=1,2,...,N$); $G$ is the universal gravitational constant; $c$ is the fundamental speed in the Minkowskian space-time; a dot between any spatial vectors, $\mathbf{a}\cdot\mathbf{b}$ denotes an Euclidean dot product of two vectors $\mathbf{a}$ and $\mathbf{b}$; $\mathrm{M}_{\tiny {B}}$ is mass of body B; $\mathbf{x}_{\tiny{B}}=\mathbf{x}_{\tiny{B}}(t)$ and $\mathbf{v}_{\tiny{B}}=\mathbf{v}_{\tiny{B}}(t)$ are coordinates and velocity of the center of mass of the body B; $\mathbf{R}_{\tiny{B}}=\mathbf{x}-\mathbf{x}_{\tiny{B}}$; $\nu_{\tiny B}$ and $\lambda_{\tiny{B}}$ are constant, but otherwise free parameters being responsible for a particular choice of the barycentric coordinates. These parameters can be chosen arbitrary for each body B of the solar system. Standard textbooks (\cite[Brumberg 1972]{vab}, \cite[Brumberg 1991]{brum}, \cite[Soffel 1989]{sof89}, and \cite[section 4.2 in Will 1993]{willbook}) assume that the coordinate parameters are equal for all bodies. This simplifies the choice of coordinates and their transformations, and allows one to identify the coordinates used by different authors. For instance, $\nu=\lambda=0$ corresponds to harmonic or isotropic coordinates (\cite[Fock 1959]{fockbook}), $\lambda=0$ and $\nu=1/2$ realizes the standard coordinates used in the book of \cite[Landau \& Lifshitz (1975)]{lali} and in PPN formalism (\cite[Will 1993]{willbook}). The case of $\nu=0, \lambda=2$ corresponds to the Gullstrand-Painlev\'e coordinates (\cite[Painlev\'e 1921]{painleve}, \cite[Gullstrand 1922]{gullstrand}), but they have not been used so far in relativistic celestial mechanics of the solar system. We prefer to have more freedom in transforming EIH equations of motion and do not equate the coordinate parameters $\nu_B$, $\lambda_B$ for different massive bodies.

If the bodies in the N-body problem are numbered by indices B, C, D, etc., and the coordinate freedom is described by equations (\ref{gt1}), (\ref{gt2}), EIH equations have the following form (\cite[equation 88 in Brumberg 1972]{vab})
\begin{eqnarray}
\label{eih1}
a^i_{\tiny{B}} & = & F^i_{\tiny{N}} + \frac1{c^2} F^i_{\tiny{EIH}}\;,
\end{eqnarray}
where the Newtonian force
\begin{eqnarray}
  \label{eih2}
  F^i_{\tiny{N}} & = & -\sum_{\tiny{C\neq B}}\frac{G\mathrm{M}_{\tiny{C}}R^i_{\tiny{BC}}}{R^3_{\tiny{BC}}}\;,
\end{eqnarray}
the post-Newtonian perturbation
\begin{eqnarray}
 \label{eih3}
 F^i_{\tiny{EIH}} & = & -\sum_{\tiny{C\neq B}}\frac{G\mathrm{M}_{\tiny{C}}R^i_{\tiny{BC}}}{R^3_{\tiny{BC}}}\Bigg\{(1+\lambda_{\tiny{C}})v^2_{\tiny{B}}-(4+2\lambda_{\tiny{C}})(\mathbf{v}_{\tiny{B}}\cdot\mathbf{v}_{\tiny C})+(2+\lambda_{\tiny{C}})v^2_{\tiny{C}}\nonumber\\
 & &-\frac{3}{2}\bigg(\frac{\mathbf{R}_{\tiny{BC}}\cdot\mathbf{v}_{\tiny{C}}}{R_{\tiny{BC}}}\bigg)^2 -3\lambda_{\tiny{C}}\bigg[\frac{\mathbf{R}_{\tiny{BC}}\cdot\mathbf{v}_{\tiny{BC}}}{R_{\tiny{BC}}}\bigg]^2-(5-2\lambda_{\tiny{B}})\frac{G\mathrm{M}_{\tiny{B}}}{R_{\tiny{BC}}}-(4-2\lambda_{\tiny{C}})\frac{G\mathrm{M}_{\tiny{C}}}{R_{{BC}}}\nonumber\\
 & &-\sum_{\tiny{D\neq B,C}}G\mathrm{M}_{\tiny D}\bigg[\frac{1}{R_{\tiny{CD}}}+\frac{4-2\lambda_{\tiny{D}}}{R_{\tiny{BD}}}-\bigg(\frac{1+2\lambda_{\tiny{C}}}{2R^3_{\tiny{CD}}}-\frac{\lambda_{\tiny{C}}}{R^3_{\tiny{BD}}}+\frac{3\lambda_{\tiny{D}}}{R_{\tiny{BD}}R^2_{\tiny{BC}}}-\frac{3\lambda_{\tiny{D}}}{R_{\tiny{CD}}R^2_{\tiny{BC}}}\bigg)\nonumber\\
 & & \times(\mathbf{R}_{\tiny{BC}}\cdot\mathbf{R}_{\tiny{CD}})\bigg]\Bigg\} -\sum_{\tiny{C\neq B}}\Bigg\{\frac{G\mathrm{M}_{\tiny C}v^i_{\tiny{CB}}}{R^3_{\tiny{BC}}}\bigg[(4-2\lambda_{\tiny{C}})(\mathbf{v}_{\tiny{B}}\cdot\mathbf{R}_{\tiny{BC}})-(3-2\lambda_{\tiny{C}})\nonumber\\
  & & \times(\mathbf{v}_{\tiny C}\cdot\mathbf{R}_{\tiny{BC}})\bigg]+\frac{G\mathrm{M}_{\tiny C}}{R_{\tiny{BC}}}\sum_{\tiny{D\neq B,C}}G\mathrm{M}_{\tiny D}R^i_{\tiny{CD}}\bigg(\frac{7-2\lambda_{\tiny{C}}}{2R^3_{\tiny{CD}}}+\frac{\lambda_{\tiny{C}}}{R^3_{\tiny{BD}}}+\frac{\lambda_{\tiny{D}}}{R_{\tiny{CD}}R^2_{\tiny{BC}}}\nonumber\\
  & & -\frac{\lambda_{\tiny{D}}}{R_{\tiny{BD}}R^2_{\tiny{BC}}}\bigg)\Bigg\}\;,
\end{eqnarray}
and $\mathbf{v}_{\tiny B}=\mathbf{v}_{\tiny B}(t)$ is velocity of the body B, $\mathbf{a}_{\tiny{B}}=\dot{\mathbf{v}}_{\tiny{B}}(t)$ is its acceleration, $\mathbf{R}_{\tiny{BC}}=\mathbf{x}_{\tiny{B}}-\mathbf{x}_{\tiny{C}}$, $\mathbf{R}_{\tiny{CD}}=\mathbf{x}_{\tiny{C}}-\mathbf{x}_{\tiny{D}}$ are relative distances between the bodies, and $\mathbf{v}_{\tiny{CB}}=\mathbf{v}_{\tiny{C}}-\mathbf{v}_{\tiny{B}}$ is a relative velocity.

Barycentric coordinates $\mathbf{x}_{\tiny{B}}$ and velocities $\mathbf{v}_{\tiny{B}}$ of the center of mass of body $B$ are adequate theoretical quantities for description of the world-line of the body with respect to the center of mass of the solar system. However, the barycentric coordinates are global coordinates covering the entire solar system. Therefore, they have little help for efficient physical decoupling of the post-Newtonian effects existing in the description of the local dynamics of the orbital motion of the Moon around Earth (\cite[Brumberg \& Kopeikin 1989]{Brumberg1989}). The problem originates from the covariant nature of EIH equations and the gauge freedom of the general relativity theory. Its resolution requires a novel approach based on introduction of a set of local coordinates associated with the barycenter of the Earth-Moon system, the Earth and the Moon (\cite[Kopeikin \& Xie 2009]{Kopeikin2009}).

The gauge freedom is already seen in the post-Newtonian EIH force (\ref{eih3}) as it explicitly depends on the choice of spatial coordinates through the gauge-fixing parameters $\lambda_{\tiny{C}}, \lambda_{\tiny{D}}$. Each term, depending explicitly on $\lambda_{C}$ and $\lambda_{D}$ in equation (\ref{eih3}), has no direct physical meaning because it can be eliminated after making a specific choice of these parameters. In many works on experimental gravity and applied astronomy (including JPL epehemrides) researches fix parameters $\lambda_{C}=\lambda_{D}=0$, which corresponds to working in harmonic coordinates. Harmonic coordinates simplify EIH equations to large extent but one has to keep in mind that they have no physical privilege anyway, and that a separate term or a limited number of terms from EIH equations of motion can not be measured if they are gauge-dependent (\cite[Brumberg 1991]{brum}). 

This opinion was recently confronted in publications by \cite[Murphy \etal\ (2007a,b)]{2007PhRvL..98g1102M,2007PhRvL..98v9002M}, \cite[Soffel \etal\ (2008)]{2008PhRvD..78b4033S}, \cite[Williams \etal\ (2004)]{test2004}, who followed \cite[Nordtvedt (1988)]{1988IJTP...27.1395N}. They separated EIH equations (\ref{eih1})-(\ref{eih3}) to the form being similar to the Lorentz force in electrodynamics
\begin{eqnarray}
  \label{eh4}
  a^i_{B} & = & \sum_{C\neq B}\bigg[E^i_{BC}+\frac{4-2\lambda_{C}}{c}(\mathbf{v}_{B}\times\mathbf{H}_{BC})^i-\frac{3-2\lambda_{C}}{c}(\mathbf{v}_{C}\times\mathbf{H}_{BC})^i\bigg]
\end{eqnarray}
where $E^i_{BC}$ is called the ``gravitoelectric'' force, and the terms associated with the cross products $(\mathbf{v}_{B}\times\mathbf{H}_{BC})^i$ and $(\mathbf{v}_{C}\times\mathbf{H}_{BC})^i$ are referred to as the ``gravitomagnetic'' force (\cite[Nordtvedt 1988]{1988IJTP...27.1395N}). The ``gravitomagnetic'' field is given by equation
\begin{eqnarray}
  \label{eh7}
  H^i_{BC} & = & -\frac{1}{c}\left(\mathbf{v}_{{BC}}\times\mathbf{E}_{BC}\right)^i= \frac{G\mathrm{M}_{C}}{c}\frac{\left(\mathbf{v}_{BC}\times\mathbf{R}_{BC}\right)^i}{R^3_{BC}}\;,
\end{eqnarray}
and is proportional to the Newtonian force multiplied by the factor of $v_{BC}/c$, where $v_{BC}$ is the relative velocity between two gravitating bodies. 

The gravitomagnetic field is of paramount importance for theoretical foundation of general relativity (\cite[Ciufolini \& Wheeler 1995]{Ciufolini1995}). Therefore, it is not surprising that the acute discussion has started about whether LLR can really measure the ``gravitomagnetic'' field $H^i_{BC}$ (\cite[Murphy \etal\ 2007a]{2007PhRvL..98g1102M}, \cite[Kopeikin 2007]{k07}, \cite[Murphy \etal\ 2007b]{2007PhRvL..98v9002M}, \cite[Ciufolini 2007]{Ciufolini2007}, \cite[Soffel \etal\ 2008]{2008PhRvD..78b4033S}). It is evident that equation (\ref{eh4}) demonstrates a strong dependence of the ``gravitomagnetic'' force of each body on the choice of the barycentric coordinates. For this reason, by changing the coordinate parameter $\lambda_{C}$ one can eliminate either the term $\left(\mathbf{v}_{B}\times\mathbf{H}_{{BC}}\right)^i$  or $\left(\mathbf{v}_{C}\times\mathbf{H}_{BC}\right)^i$ from EIH equations of motion (\ref{eh4}). In particular, the term $(\textbf{v}_B\times \textbf{H}_{BC})^i$ vanishes in the Painlev\'e coordinates, making the statement of \cite[Murphy \etal\ (2007a,b)]{2007PhRvL..98g1102M,2007PhRvL..98v9002M} about its ``measurement'' unsupported, because the strength of the factual ``gravitomagnetic'' force is coordinate-dependent. Notice that the barycentric (SSB) frame remains the same. We eliminate the ``gravitomagnetic'' force by changing the spatial coordinate only. In particular, the Lorentz transformation does not play any role. Hence a great care should be taken in order to properly interpret the LLR ``measurement'' of such gravitomagnetic terms in consistency with the covariant nature of the general theory of relativity and the theory of astronomical measurements in curved space-time. We keep up the point that the ``gravitomagnetic'' field (\ref{eh7}) is unmeasurable with LLR due to its gauge-dependence that is not associated with the transformation from one frame to another but with the coordinate transformation (\ref{gt2}).

Nevertheless, the observable LLR time delay is gauge invariant. This is because the gauge transformation changes not only the gravitational force but the solution of the equation describing the light ray propagation. For this reason, the gauge parameter $\lambda_C$ appears in the time delay \emph{explicitly}
\begin{eqnarray}
  \label{timedelay}
  t_2-t_1 & = & \frac{R_{12}}{c}+2\sum_C\frac{G\mathrm{M}_C}{c^3}\ln\bigg[\frac{R_{1C}+R_{2C}+R_{12}}{R_{1C}+R_{2C}-R_{12}}\bigg]\nonumber\\
  &&+\sum_C\lambda_C\frac{G\mathrm{M}_C}{c^3}\frac{(R_{1C}-R_{2C})^2-R_{12}^2}{2R_{1C}R_{2C}R_{12}}(R_{1C}+R_{2C}).
\end{eqnarray}
At the same time the ``Newtonian'' distance $R_{12}$ depends on the parameter $\lambda_C$ \emph{implicitly} through the solution of EIH equations (\ref{eih1})-(\ref{eih3}). This implicit dependence of the right side of (\ref{timedelay}) is exactly compensated by the explicit dependence of (\ref{timedelay}) on $\lambda_C$, making the time delay gauge-invariant.

Papers (\cite[Murphy \etal\ 2007a,b]{2007PhRvL..98g1102M,2007PhRvL..98v9002M}, \cite[Williams \etal\ 2004]{test2004}, \cite[Soffel \etal\ 2008]{2008PhRvD..78b4033S}) do not take into account the explicit gauge-dependence of the light time delay on $\lambda_C$. If the last term in (\ref{timedelay}) is omitted but EIH force is taken in form (\ref{eh4}), the equations (\ref{eh4}) and (\ref{timedelay}) become theoretically incompatible. In this setting LLR ``measures'' only the consistency of the EIH equations with the expression for time delay of the laser pulse. However, this is not a test of gravitomagnetism, which actual detection requires more precise measurement of the gauge-invariant components of the Riemann tensor associated directly either with the spin multipoles of the gravitational field of the Earth (\cite[Ciufolini 2008]{Ciufolini2008}, \cite[Ciufolini \& Pavlis 2004]{Ciufolini2004}, \cite[Ries 2009]{Ries2009}) or with the current-type multipoles of the tidal gravitational field of external bodies (\cite[Kopeikin 2008]{Kopeikin2008}).

In order to disentangle physical effects from numerous gauge dependent terms in the equations of motion of the Moon we need a precise analytic theory of reference frames in the lunar motion that includes several reference frames: SSB, GRF, SRF and EMB. This gauge-invariant approach to the lunar motion has been initiated in our paper (\cite[Kopeikin \& Xie 2009]{Kopeikin2009}) to which we refer the reader for further particular details.

\begin{acknowledgments}
Y. Xie is thankful to the Department of Physics \& Astronomy of the University of Missouri-Columbia for hospitality and accommodation. The work of Y. Xie was supported by the China Scholarship Council Grant No. 2008102243 and IAU travel grant. The work of S. Kopeikin was supported by the Research Council Grant No. C1669103 of the University of Missouri-Columbia and by 2009 faculty incentive grant of the Arts and Science Alumni Organization of the University of Missouri-Columbia.
\end{acknowledgments}

\end{document}